\documentclass[aps,prl,superscriptaddress,showpacs,amsfonts,twocolumn,floatfix]{revtex4-1}
\usepackage{graphicx}
\usepackage{amsmath}%for `split'

\usepackage{color}
\usepackage[normalem]{ulem} % \sout{old text} for strikeout
\renewcommand\sout{\bgroup \color{red} \ULdepth=-.5ex \ULset}

\begin{document}

%\preprint{}

\title{First Measurements of the Double-Polarization Observables $F$, $P$, and $H$ in 
$\omega$~Photoproduction off Transversely Polarized Protons in the $N^\ast$ Resonance Region}


%\newcommand{\comment}[1]{{\color{blue} #1}}
\newcommand{\bl}[1]{{\color{blue} #1}}
\newcommand*{\UM}{Department of Physics, University of Michigan, Ann Arbor, Michigan 48109,USA}
\newcommand*{\KAERI}{Korea Atomic Energy Research Institute, Gyeongju-si, 38180, South Korea}
\newcommand*{\NRC}{NRC ``Kurchatov Institute'', PNPI, 188300, Gatchina, Russia}

%%%%%%%%%%%%%%% Latex Macros for institute addresses  %%%%%%%%%%%%%%%%%%%%%%%%% 
\newcommand*{\ANL}{Argonne National Laboratory, Argonne, Illinois 60439, USA}
\affiliation{\ANL}
\newcommand*{\ASU}{Arizona State University, Tempe, Arizona 85287-1504, USA}
\affiliation{\ASU}
\newcommand*{\BONN}{Helmholtz-Institut f\"ur Strahlen- und Kernphysik, Universit\"at Bonn, 53115 Bonn, Germany}
\affiliation{\BONN}
\newcommand*{\CANISIUS}{Canisius College, Buffalo, New York 14208, USA}
\affiliation{\CANISIUS}
\newcommand*{\CMU}{Carnegie Mellon University, Pittsburgh, Pennsylvania 15213, USA}
\affiliation{\CMU}
\newcommand*{\CUA}{Catholic University of America, Washington, D.C. 20064, USA}
\affiliation{\CUA}
\newcommand*{\SACLAY}{IRFU, CEA, Universit\'e Paris-Saclay, F-91191 Gif-sur-Yvette, France}
\affiliation{\SACLAY}
\newcommand*{\CNU}{Christopher Newport University, Newport News, Virginia 23606, USA}
\affiliation{\CNU}
\newcommand*{\UCONN}{University of Connecticut, Storrs, Connecticut 06269, USA}
\affiliation{\UCONN}
\newcommand*{\DUKE}{Duke University, Durham, North Carolina 27708-0305, USA}
\affiliation{\DUKE}
\newcommand*{\FU}{Fairfield University, Fairfield CT 06824, USA}
\affiliation{\FU}
\newcommand*{\FERRARAU}{Universit\`a di Ferrara, 44121 Ferrara, Italy}
\affiliation{\FERRARAU}
\newcommand*{\FIU}{Florida International University, Miami, Florida 33199, USA}
\affiliation{\FIU}
\newcommand*{\FSU}{Florida State University, Tallahassee, Florida 32306, USA}
\affiliation{\FSU}
\newcommand*{\Genova}{Universit\`a di Genova, 16146 Genova, Italy}
\affiliation{\Genova}
\newcommand*{\GWUI}{The George Washington University, Washington, DC 20052, USA}
\affiliation{\GWUI}
\newcommand*{\UG}{University of Georgia, Athens, GA30602, USA}
\affiliation{\UG}
\newcommand*{\ISU}{Idaho State University, Pocatello, Idaho 83209, USA}
\affiliation{\ISU}
\newcommand*{\INFNFE}{INFN, Sezione di Ferrara, 44100 Ferrara, Italy}
\affiliation{\INFNFE}
\newcommand*{\INFNFR}{INFN, Laboratori Nazionali di Frascati, 00044 Frascati, Italy}
\affiliation{\INFNFR}
\newcommand*{\INFNGE}{INFN, Sezione di Genova, 16146 Genova, Italy}
\affiliation{\INFNGE}
\newcommand*{\INFNRO}{INFN, Sezione di Roma Tor Vergata, 00133 Rome, Italy}
\affiliation{\INFNRO}
\newcommand*{\INFNTUR}{INFN, Sezione di Torino, 10125 Torino, Italy}
\affiliation{\INFNTUR}
\newcommand*{\ORSAY}{Institut de Physique Nucl\'eaire, IN2P3-CNRS, Universit\'e Paris-Sud, Universit\'e Paris-Saclay, F-91406 Orsay, France}
\affiliation{\ORSAY}
\newcommand*{\Juelich}{Institute f\"ur Kernphysik, 52425 J\"ulich, Germany}
\affiliation{\Juelich}
\newcommand*{\ITEP}{Institute of Theoretical and Experimental Physics, Moscow, 117259, Russia}
\affiliation{\ITEP}
\newcommand*{\JMU}{James Madison University, Harrisonburg, Virginia 22807, USA}
\affiliation{\JMU}
\newcommand*{\KNU}{Kyungpook National University, Daegu 41566, Republic of Korea}
\affiliation{\KNU}
\newcommand*{\LAMAR}{Lamar University, 4400 MLK Blvd, PO Box 10009, Beaumont, Texas 77710, USA}
\affiliation{\LAMAR}
\newcommand*{\MISS}{Mississippi State University, Mississippi State, MS 39762-5167, USA}
\affiliation{\MISS}
\newcommand*{\UNH}{University of New Hampshire, Durham, New Hampshire 03824-3568, USA}
\affiliation{\UNH}
\newcommand*{\NSU}{Norfolk State University, Norfolk, Virginia 23504, USA}
\affiliation{\NSU}
\newcommand*{\OHIOU}{Ohio University, Athens, Ohio 45701, USA}
\affiliation{\OHIOU}
\newcommand*{\ODU}{Old Dominion University, Norfolk, Virginia 23529, USA}
\affiliation{\ODU}
\newcommand*{\URICH}{University of Richmond, Richmond, Virginia 23173, USA}
\affiliation{\URICH}
\newcommand*{\ROMAII}{Universit\`a di Roma Tor Vergata, 00133 Rome, Italy}
\affiliation{\ROMAII}
\newcommand*{\MSU}{Skobeltsyn Institute of Nuclear Physics, Lomonosov Moscow State University, 119234 Moscow, Russia}
\affiliation{\MSU}
\newcommand*{\SCAROLINA}{University of South Carolina, Columbia, South Carolina 29208, USA}
\affiliation{\SCAROLINA}
\newcommand*{\TEMPLE}{Temple University, Philadelphia, PA 19122, USA}
\affiliation{\TEMPLE}
\newcommand*{\JLAB}{Thomas Jefferson National Accelerator Facility, Newport News, Virginia 23606, USA}
\affiliation{\JLAB}
\newcommand*{\UTFSM}{Universidad T\'{e}cnica Federico Santa Mar\'{i}a, Casilla 110-V Valpara\'{i}so, Chile}
\affiliation{\UTFSM}
\newcommand*{\GLASGOW}{University of Glasgow, Glasgow G12 8QQ, United Kingdom}
\affiliation{\GLASGOW}
\newcommand*{\YORK}{University of York, York YO10, United Kingdom}
\affiliation{\YORK}
\newcommand*{\VT}{Virginia Tech, Blacksburg, Virginia 24061-0435, USA}
\affiliation{\VT}
\newcommand*{\VIRGINIA}{University of Virginia, Charlottesville, Virginia 22901, USA}
\affiliation{\VIRGINIA}
\newcommand*{\WM}{College of William and Mary, Williamsburg, Virginia 23187-8795, USA}
\affiliation{\WM}
\newcommand*{\YEREVAN}{Yerevan Physics Institute, 375036 Yerevan, Armenia}
\affiliation{\YEREVAN}
\newcommand*{\ZHENGZHOU}{Zhengzhou University, Zhengzhou, Henan 450001, China}
\affiliation{\ZHENGZHOU}

\newcommand*{\NOWMISS}{Mississippi State University, Mississippi State, MS 39762-5167, USA}
\newcommand*{\NOWGWUI}{The George Washington University, Washington, DC 20052, USA}
\newcommand*{\NOWISU}{Idaho State University, Pocatello, Idaho 83209, USA}
\newcommand*{\NOWORNL}{Oak Ridge National Laboratory, Oak Ridge, TN 37831, USA}
\newcommand*{\NOWINFNGE}{INFN, Sezione di Genova, 16146 Genova, Italy}
\newcommand*{\SAUDI}{Imam Abdulrahman Bin Faisal University, Industrial Jubail 31961, Saudi Arabia}
 %%%%%%%%%%%%%%% END OF Latex Macros for institute addresses  %%%%%%%%%%%%%%%%%%%%%%%%%


\author{P.~Roy} \altaffiliation[Present address: ]{\UM} \affiliation{\FSU}
\author{S.~Park} \altaffiliation[Present address: ]{\KAERI} \affiliation{\FSU} 
%\author{Z.~Akbar} \affiliation{\FSU}
\author{V.~Crede} \altaffiliation[Corresponding author: ]{crede@fsu.edu} \affiliation{\FSU}
\author{A.~V.~Anisovich} \affiliation{\BONN} \affiliation{\NRC}
\author{E.~Klempt} \affiliation{\BONN}
\author{V.~A.~Nikonov} \affiliation{\BONN} \affiliation{\NRC}
\author{A.~V.~Sarantsev} \affiliation{\BONN} \affiliation{\NRC}
\author{N.~C.~Wei} \affiliation{\ZHENGZHOU}\affiliation{School of Nuclear Science and Technology, University of Chinese Academy of Sciences, Beijing 100049, China}
\author{F.~Huang} \affiliation{School of Nuclear Science and Technology, University of Chinese Academy of Sciences, Beijing 100049, China}
\author{K.~Nakayama} \affiliation{\UG}

% ...

%\thanks{}

\collaboration{CLAS Collaboration at Jefferson Lab}\noaffiliation

\date{Received: \today / Revised version:}

\begin{abstract}
First measurements of double-polarization observables in $\omega$~photoproduction off 
the proton are presented using transverse target polarization and data from the CEBAF Large 
Acceptance Spectrometer (CLAS) FROST experiment at Jefferson Lab. The beam-target asymmetry~$F$ 
has been measured using circularly polarized, tagged photons in the energy range 1200--2700~MeV, 
and the beam-target asymmetries~$H$ and~$P$ have been measured using linearly polarized, tagged 
photons in the energy range 1200--2000~MeV. %These measurements contribute to the completion 
%of the set of polarization observables which are accessible at current experimental facilities. 
These measurements significantly increase the database on polarization observables. The results 
are included in two partial-wave analyses and reveal significant contributions from several 
nucleon ($N^\ast$) resonances. In particular, contributions from new $N^\ast$~resonances listed 
in the Review of Particle Properties are observed, which aid in reaching the goal of mapping out 
the nucleon resonance spectrum.
\end{abstract}

%\keywords{}

\maketitle


Photoproduction of the isoscalar vector mesons $\omega$ and $\phi$ off the proton plays an 
important role in our understanding of many hadronic physics phenomena in the non-perturbative 
regime. Photoproduction of an $\omega$~meson at lower energies provides unique information on 
the mechanism of nucleon resonance excitation and on the strength of the $\omega N N^\ast$ 
coupling, which aids in shedding light on the structure of baryon resonances. 

The study of $\omega$-meson photoproduction is particularly interesting in the search for
new, hitherto unknown nucleon resonances. The reaction threshold lies above the thresholds for 
$\pi$ and $\eta$~photoproduction and therefore, $\omega$~photoproduction probes the higher-mass 
nucleon states above $W = 1700$~MeV. At these center-of-mass energies, the $\pi N$ and $\eta N$ 
photoproduction cross sections are significantly smaller. Moreover, the $\omega$~meson is an 
isoscalar particle and is sensitive only to $I = 1/2$~(nucleon) resonances which reduces the 
complexity of the contributing intermediate states. A discussion of recent progress toward 
understanding the nucleon resonance spectrum can be found in recent reviews, 
e.g. Refs.~\cite{Klempt:2009pi,Crede:2013sze}.

In this letter, we report on the first measurements of the polarization observables $F$, $P$, 
and $H$ for the reaction
\begin{equation}
\vec{\gamma}\,\vec{p} \to p\,\omega\quad\text{where}~\omega \to \pi^+\,\pi^-\,\pi^0\,,
\end{equation}
using linearly as well as circularly polarized tagged photons and transversely polarized protons. 
Without measuring any recoil polarization, the differential cross section for this combination is 
given by~\cite{Barker:1975bp,Fasano:1992es,Pichowsky:1994gh}
\begin{equation}
 \begin{split}
  \frac{d\sigma}{d\Omega}\,=\,\frac{d\sigma_0}{d\Omega}\,\{\,(\,1\,& -\,\delta_{\,l}\,\Sigma\, 
                           {\rm cos}\,2\beta\,)\\[0.1ex]
    & + \Lambda\,{\rm cos}\,\alpha\, ( - \delta_{\,l}\,H\,{\rm sin}\, 2 \beta\, 
                 +\, \delta_{\odot}\,F\,)\\[1ex]
    & -\, \Lambda\,{\rm sin}\,\alpha\, ( - T\, +\, 
                 \delta_{\,l}\,P\,{\rm cos}\,2 \beta\,)\,\}\,, 
\end{split}
\label{Equation:Reaction}
\end{equation}
where $\delta_{\,l}$ ($\delta_{\,\odot}$) denotes the degree of linear (circular) beam polarization 
and $\Lambda$~denotes the degree of target polarization. For transverse target polarization, the 
available polarization observables are the target asymmetry~$T$, the beam-target asymmetry $F$ using 
a circularly polarized beam, and the beam-target asymmetries $H$ and $P$ using a linearly polarized 
beam. The angle $\beta$ ($\alpha$) describes the inclination of the linear-beam (transverse-target) 
polarization with respect to the center-of-mass plane spanned by the beam axis and the recoil proton.

The FROzen-Spin Target (FROST) experiment, conducted at the Thomas Jefferson National Accelerator 
Facility, was designed to perform measurements with polarized beams and targets. The details of 
the experiment are discussed in Refs.~\cite{Akbar:2017uuk,Roy:2017qwv,Keith:2012ad}.

The CEBAF accelerator facility at Jefferson Lab delivered longitudinally polarized electrons with 
energies up to 2.4~GeV and a polarization of about 87\,\%~\cite{Roy:Thesis}. Circularly polarized 
photons were then obtained by transferring the polarization from the electrons to the photons in a 
brems\-strahlung process when the electrons scattered off an amorphous gold radiator. The larger the 
fractional energy carried by the photon with respect to the electron energy, the greater the degree of 
polarization~\cite{Akbar:2017uuk,Olsen:1959zz}.

%The degree of the photon-beam polarization, $\delta_{\,\odot}$, was given as a function of the 
%degree of electron-beam polarization, $\delta_{\,e^-}$, and the photon-beam energy, $E_{\,\gamma}$:
%\begin{equation}
%\delta_{\,\odot}\,=\,\delta_{e^-}\,.\,\frac{4x\,-\,x^2}{4\,-\,4x\,+\,3x^2}~,
%\end{equation} 
%where $x = E_{\gamma}/E_{\, e^-}$ and $E_{\, e^-}$ is the accelerator energy.

Linearly polarized photons were created via coherent brems\-strahlung by scattering unpolarized 
electrons off a diamond crystal. These polarized photons typically covered a 200-MeV-wide energy 
range below the sharp coherent edge. Data were recorded with the position of the coherent edge 
ranging from 700~MeV to 2100~MeV, in steps of 200~MeV. The degree of linear polarization was 
determined by fitting the energy distributions of the incident photons and was observed to 
vary between 40--60\,\%. The polarized photons were energy and time tagged with resolutions of 
0.1\,\% and 100~ps, respectively, using a photon tagging system~\cite{Sober:2000we}.

A state-of-the-art component of the experiment was the polarized target, described in detail in
Ref.~\cite{Keith:2012ad}. It was placed at the center of the CLAS spectrometer, and provided an
average degree of polarization of 81\,\%. %with the plane inclined at $116.1^\circ$ ($-63.9^\circ$) 
%to the horizontal when the proton polarization was pointing away from (towards) the floor of the 
%experimental hall. 
The direction of the polarization was reversed every 5--7 days.
%TEMPO-doped frozen butanol beads (${\rm C}_4{\rm H}_9{\rm OH}$) placed in a $6$~cm long container 
%served as the target material. The nuclei of the covalently bonded hydrogen atoms were polarized 
%by dynamic nuclear polarization~\cite{DNP} which involved applying a $5$~T polarizing field 
%while the target was maintained at 250\,-\,300~mK using a dilution refrigerator. The direction 
%of the target polarization was reversed typically every 4\,-\,7 days by tuning the microwave 
%frequency above or below the electron spin resonance frequency. Once the desired polarization 
%was achieved, the polarizing field was ramped down and the target was cooled to 60~mK. A dipole 
%magnet with a field strength of 0.5~T was energized to maintain the transverse polarization during 
%data-taking. The transverse-target polarization was inclined at $116.1^\circ$ ($-63.9^\circ$) with 
%respect to x$_{\rm lab}$ when pointing away (towards) the floor of the experimental hall. A high 
%degree of target polarization ($\sim\,81\%$ on average) was achieved, along with a long relaxation 
%time of 4000 (3400) hours without (with) beam. 
To study background originating from unpolarized protons of the carbon and oxygen atoms in the 
butanol target, carbon and polyethylene disks were placed at approximately 9~cm and 16~cm 
downstream of the butanol target. The vertex distribution shows distinct peaks from each target, 
allowing for a clean separation of events.

\begin{figure}
\includegraphics[width=0.47\textwidth]{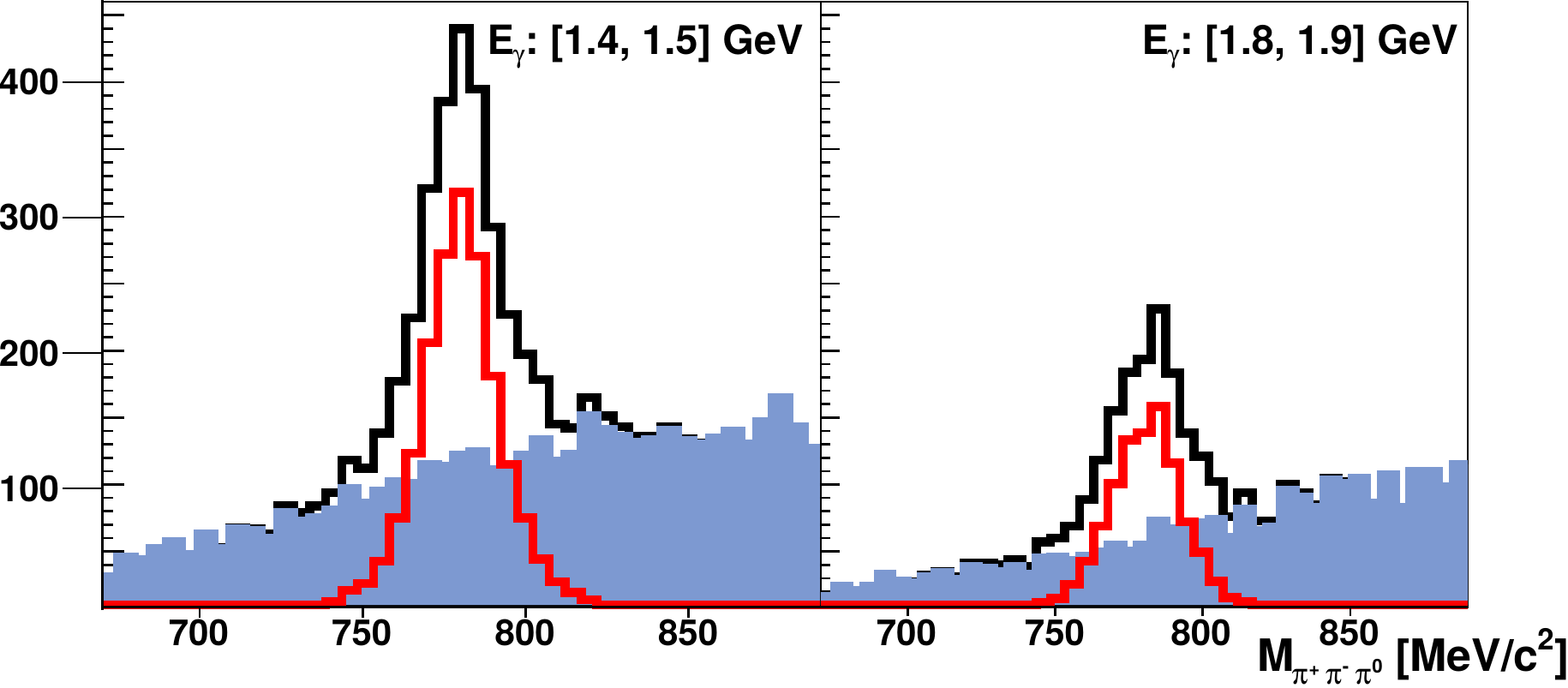}
\caption{\label{Figure:Masses} (Color online) Typical examples of signal and background mass
distributions from butanol data after applying all kinematic cuts and corrections. The invariant
$\pi^+\pi^-\pi^0$ masses are shown for two energies at the same angle,
$0.2\,<\,{\rm cos}\,\Theta^{\,\omega}_{\rm \,c.m.}\, < 0.4$. The {\bf black line} shows the unweighted
distribution from the butanol target, the {\color{blue} blue-shaded area} shows the background mass
distribution (weighted by $1-Q$), and the {\color{red} red line} shows the signal distribution (weighted
by $Q$.)}
\end{figure}

%The polarized target was placed at the center of the CLAS spectrometer. 
The CLAS detector, with its six-fold symmetry about the beamline, was capable of detecting charged 
particles with a laboratory polar-angle coverage of $[8,\,142]^{\circ}$ and almost $2\pi$ coverage in 
the azimuthal angle. The final-state particles traversed several layers of sub-detectors, including 
drift chambers (DC) and time-of-flight (TOF) scintillators. A start counter (SC) provided the initial 
time information of the events. Full details of the CLAS detector are provided in Ref.~\cite{Mecking:2003zu}. 
For an event to be recorded, the trigger conditions required at least one charged particle in the final state.

In this analysis, the~$\omega$ was reconstructed from its $\pi^+\pi^-\pi^0$~decay, which has the 
highest branching ratio (89\,\%) among all $\omega$~decay modes. Events were selected to have 
exactly one incident-photon candidate with a timing (using the photon tagger) at the event vertex 
within 1~ns of the event time provided by the SC. Only those events that had exactly one proton,
plus one positively charged and one negatively charged pion track in the final state were
retained. To further improve the particle identification, each final-state particle's $\beta$~value 
was calculated separately from its measured momentum using the DC, $\beta_{\rm \,DC}$, and from its 
measured velocity using the TOF system and the SC, $\beta_{\rm \,TOF}$. Events were selected
based on good agreement of $\beta_{\rm \,DC}$ and $\beta_{\rm \,TOF}$~\cite{Roy:2017qwv,Roy:Thesis}.
%Only those events were considered if
%\begin{equation}
% \Delta\beta \,=\,\beta_{\rm \,DC} - \beta_{\rm \,TOF}\,<\,3\sigma\,,
%\label{equ:beta}
%\end{equation}
%where $\sigma$ was the standard deviation of the $\Delta\beta$ Gaussian distribution centered at zero. 
The momenta of the final-state particles were corrected for energy losses using standard CLAS 
techniques. Additional corrections of a few MeV were required for the momentum magnitudes, which are 
discussed in detail in Refs.~\cite{Akbar:2017uuk,Roy:2017qwv,Roy:Thesis}.

A four-constraint (4C) kinematic fit to the exclusive $\gamma p \to p\,\pi^+\pi^-$ reaction imposing
energy and momentum conservation aided in tuning the full covariance matrix. The reaction 
$\gamma p \to p\,\pi^+\pi^-\,({\rm missing}\,\pi^0)$ was next kinematically fit, and events with 
a confidence level below $0.001$ were rejected, removing most of the $\pi^+\pi^-$~background. The 
remaining background consisted mostly of $p\,\omega$~events originating from unpolarized bound protons 
in the butanol (C$_4$H$_9$OH) target and non-$p\,\omega$ events resulting in a $p\,\pi^+\pi^-\pi^0$ 
final state. %These background events were removed using an 
%event-based technique~\cite{Williams:2008sh}. The details of implementing this technique for the 
%FROST data are discussed in Ref.~\cite{Roy:2017qwv}. Each event was assigned a weight (or $Q$~factor) 
%which describes the probability for the event to be a signal event. 
These were accounted for using the $Q$-factor technique, which determines the probability for an event 
to be a signal event (as opposed to background) on the basis of a sample of its nearest kinematic
neighbors in a very small region of the multi-dimensional $\pi^+\pi^-\pi^0$ phase space around the
candidate event~\cite{Roy:2017qwv,Williams:2008sh}. The method assumes that the signal and background
distributions do not vary rapidly in the selected region. The $\pi^+\pi^-\pi^0$~mass distribution of 
each event and its nearest kinematic neighbors was fit using a Voigtian for the signal probability 
function (pdf) and a third-order Chebychev polynomial for the background pdf. The value of $Q$ is 
then defined as the ratio of signal amplitude to total amplitude at the mass of the candidate
event. Figure~\ref{Figure:Masses} shows examples of signal and background distributions in the 
invariant $\pi^+\pi^-\pi^0$ mass obtained by weighting each event with $Q$ and $1-Q$, respectively.

For each bin in incident-photon energy and meson center-of-mass angle 
($E_{\gamma}$,\,cos$\,\Theta^{\,\omega}_{\rm \,c.m.}$), an event-based maximum-likelihood 
technique was applied to fit the azimuthal angular distributions of the recoil proton in the 
lab frame to extract the polarization observables~\cite{Paterson:2016vmc}. The likelihood 
function in each kinematic bin is
\begin{equation}
-{\rm ln}\,L = -\sum_{i=1}^{N_{\rm \,events}}~w_i\,{\rm ln}\,(P_{\,i})~,\quad 
 P_{\,i}\,=\, \frac{1\,\pm\,A}{2}~,
%{\cal{L}} = \prod_{i=1}^{N_{\rm \,total}} P_{{\rm event}\,i}^{\,w_i}~,\quad {\rm where}~P_{{\rm 
%  event}\,i}\,=\, \frac{1\,\pm\,A}{2}~,
\label{Equation:L}
\end{equation}
and $A = (N_1 - N_2) / (N_1 + N_2)$ denotes the asymmetry in the azimuthal angular distributions 
of events with different orientations of the beam-target polarization. The sign of $A$ depends on 
the corresponding relative orientation of the beam-target polarization in the $i$th event. The 
weights, $w_i$, depend on the $Q$~factors and additional normalization factors. More details and 
a complete list of definitions are given in Refs.~\cite{Roy:2017qwv,Roy:Thesis}.

The asymmetry~$A$ depends on the differential cross section (Eqn.~\ref{Equation:Reaction}) and 
hence, on the polarization observables. Maximizing the likelihood ${\cal{L}}$ gives the most likely 
values for the observables. Owing to statistical limitations, a simultaneous fit to all polarization 
observables did not converge. Different data sets, corresponding to the different orientations of the 
beam-target polarization, were combined with appropriate normalization factors to reduce the number 
of unknown parameters in the likelihood expression. The observable $F$ was determined separately
using circular beam polarization, whereas the observables $H$ and $P$ were determined from simultaneous 
fits using linear beam polarization (see Eqn.~\ref{Equation:Reaction}).

A major contribution to the overall systematic uncertainty came from the background subtraction. 
This $Q$~factor uncertainty was determined for all observables in each ($E_{\gamma}$,\,cos\,
$\Theta^{\,\omega}_{\rm \,c.m.}$)~bin by modifying each $Q$~factor by its corresponding fit 
uncertainty $\sigma_{Q}$, and re-extracting the observable. The absolute difference was taken 
as the systematic uncertainty and averaged about 8\,\% for incident-photon energies $>1.3$~GeV. 
Other sources of uncertainty included the degree of linear- (circular-)\,beam polarization 
($5\,\%$~($4\,\%$)), the degree of transverse-target polarization ($2\,\%$), the direction of the 
target polarization ($2\,\%$) and the flux normalization. The latter was $5\,\%$ for data with 
linear-beam polarization, and $2\,\%$ for data with circular-beam polarization since the beam 
helicity flipped rapidly leading to the same photon flux for opposite beam helicities. Gray 
%bands in Figs.~\ref{Figure:F-Observable} \&~\ref{Figure:P-H-Observables} show only absolute 
bands in the figures show only absolute systematic uncertainties due to the background subtraction; 
scale-type uncertainties are not included.

\begin{figure}
\includegraphics[width=0.5\textwidth,height=0.47\textheight]{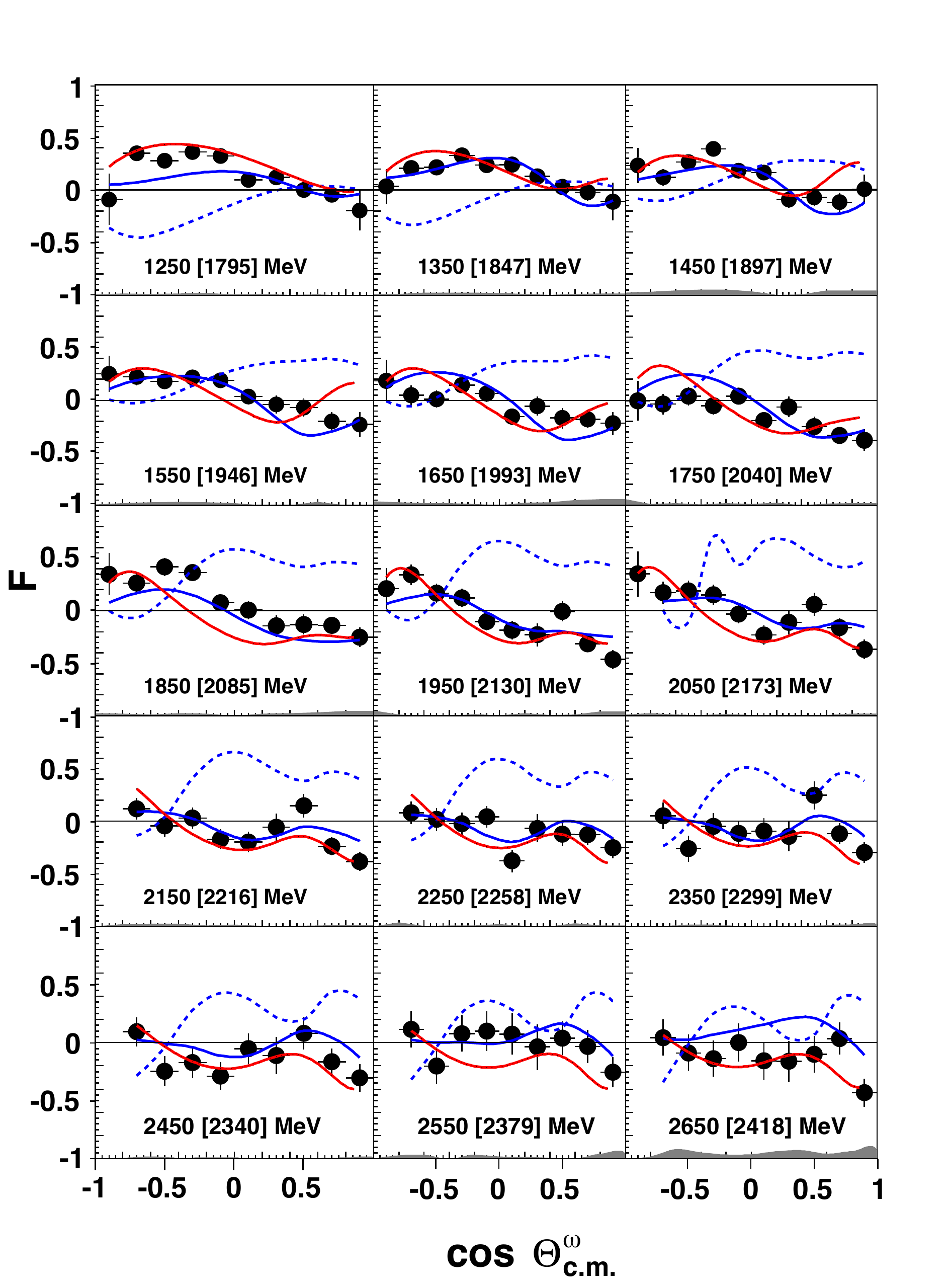}
\caption{\label{Figure:F-Observable} First-time measurement of the beam-target asymmetry $F$ in 
$\vec{\gamma}\,\vec{p} \to p\,\omega$. Shown are distributions in 100-MeV-wide incident-photon energy 
bins (labeled as $E_\gamma\,[W]$\,) as a function of cos\,$\Theta^{\,\omega}_{\rm \,c.m.}$ in the 
center-of-mass frame. Each data point has been assigned its statistical uncertainty, whereas the gray 
band at the bottom of each panel represents the absolute systematic uncertainties due to the background 
subtraction. The blue and red solid curves show the BnGa PWA solution and fits by Wei {\it et al.}
\cite{Wei:2018}, respectively. The blue dashed curve denotes an earlier BnGa solution~\cite{Denisenko:2016ugz}.}
\end{figure}

\begin{figure*}
\includegraphics[width=1.0\textwidth]{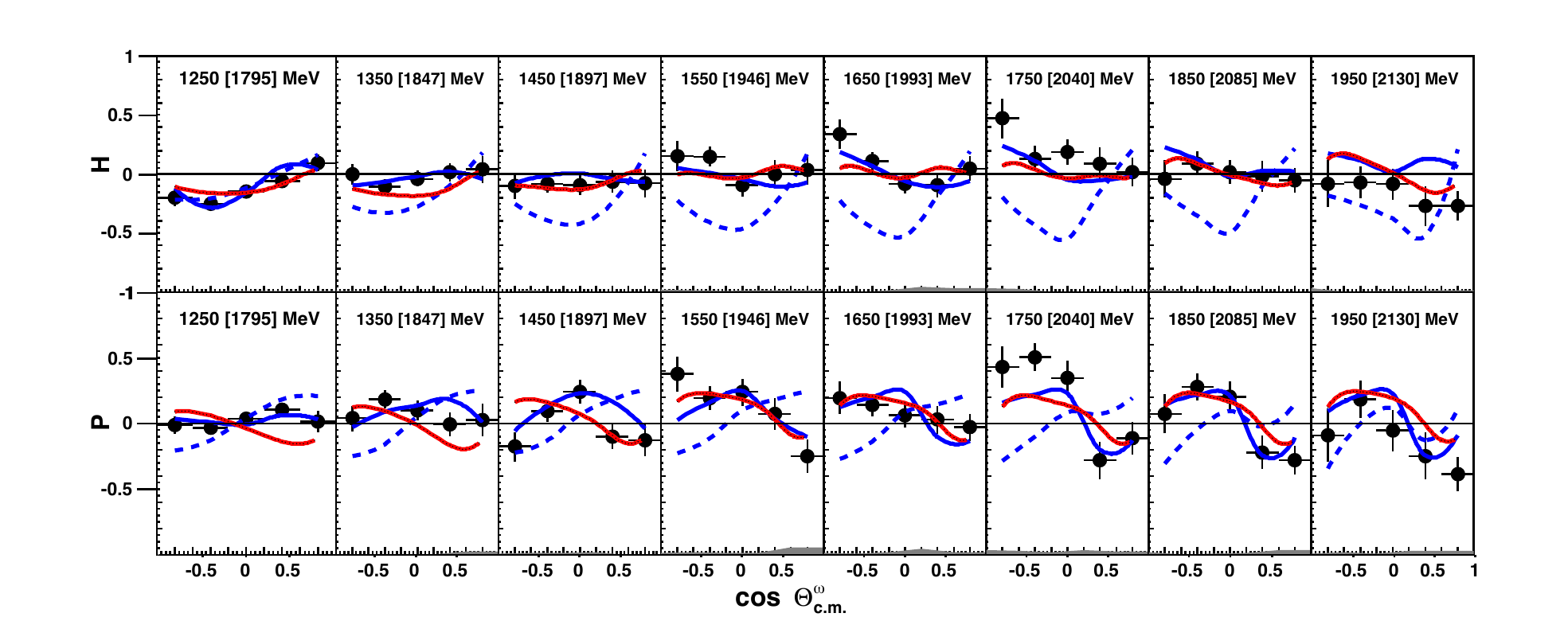}
\caption{\label{Figure:P-H-Observables} (Color online) First-time measurements of the beam-target 
asymmetries $H$ (top) and $P$ (bottom) in $\omega$~photoproduction off the proton. Shown are eight 
$100$-MeV-wide energy bins (labeled as $E_\gamma\,[W]\,$) as a function of 
cos\,$\Theta^{\,\omega}_{\rm \,c.m.}$ in the center-of-mass frame. Each data point has been assigned 
its statistical uncertainty, whereas the gray band at the bottom of each panel represents the absolute 
systematic uncertainties due to the background subtraction. The blue and red solid curves show the 
BnGa PWA solution and fits by Wei {\it et al.}~\cite{Wei:2018}, respectively. The blue dashed curve 
denotes an earlier BnGa solution~\cite{Denisenko:2016ugz}.}
\end{figure*}

The $\omega$~polarization observables presented here are first-time measurements, representing a
substantial increase in the world database for $\omega$~photoproduction. Figure~\ref{Figure:F-Observable}
shows the beam-target asymmetry $F$ and Fig.~\ref{Figure:P-H-Observables} shows the beam-target
asymmetries $H$ and $P$ for the incident-photon energy range 1200--2000~MeV in 100-MeV-wide bins
and 10 and 5~cos$\,\Theta^{\,\omega}_{\rm \,c.m.}$~bins in the center-of-mass frame, respectively.
The asymmetries are substantial and vary significantly with energy, indicative of strong contributions
from nucleon resonances.

The role of $N^\ast$~resonances in $\omega$~photoproduction has long been discussed in the literature, 
e.\,g., using effective Lagrangian~\cite{Oh:2000zi,Zhao:2000tb,Titov:2002iv} and coupled-channel 
K-matrix approaches~\cite{Penner:2002md,Shklyar:2004ba}. Given the scarcity of data at the time, 
most of these studies were based only on the differential cross section data, and not surprisingly 
disagree on the contribution of $N^\ast$~resonances.

The data presented here, and further $\omega$~data from the FROST experiment on the helicity 
asymmetry $E$~\cite{Akbar:2017uuk} and on the single-polarization observables 
$\Sigma$~\cite{Collins:2017vev,Roy:2017qwv} and T~\cite{Roy:2017qwv} (beam and target asymmetries, 
respectively) were included in two independent analyses: A partial-wave analysis (PWA) within the 
Bonn-Gatchina (BnGa) coupled-channel framework~\cite{Anisovich:2006bc} based on a large database of 
pion- and photon-induced reactions~\cite{BnGa:Database}, and a tree-level-based effective Lagrangian 
approach~\cite{Wei:2018}, shown in Figs.~\ref{Figure:F-Observable} and~\ref{Figure:P-H-Observables} 
as the solid and dashed lines, respectively. In contrast to the coupled, multi-channel BnGa
analysis, the effective Lagrangian approach of Ref.~\cite{Wei:2018} considers only the $\omega N$
channel. The reaction amplitude consists of $s$-, $t$-, and $u$-channel Feynman diagrams combined
with a phenomenological contact current which accounts for effects not explicitly included and is 
required for local gauge invariance of the overall amplitude. More details are given in 
Refs.~\cite{Wang:2017tpe,Wang:2018vlv}.

The BnGa description of these new data started with a PWA solution of an earlier analysis that is 
discussed in Ref.~\cite{Denisenko:2016ugz}. This initial analysis was based on results in 
$\gamma p\to p\,\omega~(\omega\to\pi^0\gamma)$ obtained by the CBELSA/TAPS Collaboration on 
differential cross sections~\cite{Wilson:2015uoa}, the double-polarization observables {\bf G}, 
{\bf G$_{\pi}$}~\cite{Eberhardt:2015lwa}, the beam asymmetry $\Sigma$~\cite{Klein:2008aa} and a 
variety of spin-density matrix elements (SDMEs): $\rho^{1}_{00}$, $\rho^{1}_{11}$, $\rho^{1}_{1-1}$, 
$\rho^{1}_{10}$, $\rho^{2}_{10}$, $\rho^{2}_{1-1}$ (using linear-beam polarization) as well as 
$\rho^{0}_{00}$, $\rho^{0}_{10}$, $\rho^{0}_{1-1}$ (unpolarized beam)~\cite{Wilson:2015uoa}.

The earlier analysis revealed significant $t$-channel contributions from the exchange of pomerons, 
which increase with energy and account for about 50\,\% of the total cross section at about $W=2$~GeV. 
Moreover, the polarization observables and SDMEs revealed notable contributions from as many as 
12~nucleon resonances, and several branching ratios were determined for the first 
time~\cite{Denisenko:2016ugz}. Evidence was found for the poorly known states $N(1880)\,1/2^+$, 
$N(2000)\,5/2^+$, $N(1895)\,1/2^-$, and $N(2120)\,3/2^-$. Small contributions were also revealed 
from several weaker partial waves. However, this solution provided a poor description of the new 
CLAS polarization observables (see Figs.~\ref{Figure:F-Observable} and \ref{Figure:P-H-Observables}): 
$F$, $H$, $P$, and $T$. Particularly, the predicted target asymmetry appeared to have the wrong sign 
using the definitions for these observables from Ref.~\cite{Fasano:1992es}.

The BnGa~solution for the new CLAS data presented here confirms the five dominant partial wave 
amplitudes that were reported in Ref.~\cite{Denisenko:2016ugz}. The $J^P = 3/2^+$~partial wave 
exhibits a significant peak close to $W=1800$~MeV that is identified with the $N(1720)\,3/2^+$
resonance. A notable contribution from the $3/2^-$~partial wave is observed above 2~GeV and 
identified with the $N(2120)\,3/2^-$. Compared with earlier findings, the coupling of the 
$N(1875)\,3/2^-$ to $N\omega$ has decreased by about $70\,\%$. The intensity appears to have 
shifted to the $5/2^+$~partial wave above $W = 1900$~MeV, where the contribution of the 
$N(2000)\,5/2^+$~state has been observed to increase by about 50\,\%. The $1/2^-$~partial wave 
exhibits a smoother behavior, but the analysis found that the coupling to $N(1895)\,1/2^-$ has 
not significantly changed. This smoother behavior is a result of a sign change in the contribution 
of the non-resonant amplitudes. The dominant contributions, in particular of the $N(2000)\,5/2^+$
state, are consistent with the results of a single-channel PWA by the CLAS Collaboration~\cite{Williams:2009aa}.

The effective Lagrangian approach by Wei {\it et al.}~\cite{Wei:2018} is based on all published 
data from the CLAS Collaboration, including the new double-polarization observables discussed here. 
To achieve a good description of the data, seven nucleon resonances have been added in the analysis. 
A significant peak in the $3/2^+$~wave around $W=1800$~MeV is confirmed, which originates from the 
$N(1720)\,3/2^+$. The $3/2^-$~partial wave shows important contributions, which mainly stem from 
the $N(1520)\,3/2^-$ and $N(1700)\,3/2^-$ resonances ($W<2$~GeV), in agreement with the BnGa~analysis. 
The latter two resonances prove to be important in the description of the new $F$ and $H$~observables.
This analysis also identifies significant contributions from the $5/2^+$~partial wave, again consistent
with the findings of the BnGa group.

In summary, the beam-target double-polarization observables $F$, $P$, and $H$ in the reaction 
$\vec{\gamma}\,\vec{p} \to p\,\omega$ have been measured for the first time across the $N^\ast$
resonance region. Convergence among different groups on the leading $N^\ast$~resonance contributions 
appears imminent based on these new measurements. Several poorly known states have been identified in 
$\omega$~photoproduction. Particularly noteworthy are contributions from the new $N^\ast$~states that
have been listed in the Review of Particle Properties since 2014 based on photoproduction experiments.
In the $3/2^-$~partial wave for example, contributions from the recently added $N(1875)\,3/2^-$ 
and $N(2120)\,3/2^-$ states are observed. Also identified in $\omega$~photoproduction is the new 
$N(1880)\,1/2^+$~state which, together with the $N(1900)\,3/2^+$ and $N(1990)\,7/2^+$ states, and 
the poorly established $N(2000)\,5/2^+$~state, is considered to form a quartet of nucleon states 
in the $({\bf 70},\,2^+_2)$~supermultiplet with quark spin $S = 3/2$ and positive parity. Some 
open questions remain, including the relative strength of $t$-channel contributions close to the 
reaction threshold from the exchange of either pomerons or pions. A full discussion of the contributing 
$N^\ast$~resonances, their $N\omega$~couplings, and the impact of particular observables will be available 
in forthcoming papers~\cite{Anisovich:2018,Wei:2018}.

The authors gratefully acknowledge the excellent support of the technical staff at Jefferson 
Lab and all participating institutions. This research is based on work supported by the U.\,S. 
Department of Energy, Office of Science, Office of Nuclear Physics, under Contract No. 
DE-AC05-06OR23177. The group at Florida State University acknowledges additional support from 
the U.S. Department of Energy, Office of Science, Office of Nuclear Physics, under Contract No. 
DE-FG02-92ER40735. This work was also supported by the U.\,S. National Science Foundation, the 
State Committee of Science of Republic of Armenia, the Chilean Comisi\'on Nacional de Investigaci\'on 
Cientifica y Tecnol\'ogica, the Italian Istituto Nazionale di Fisica Nucleare, the French Centre 
National de la Recherche Scientifique, the French Commissariat a l’Energie Atomique, the Scottish 
Universities Physics Alliance (SUPA), the United Kingdom Science and Technology Facilities
Council (STFC), the National Research Foundation of Korea, the Deutsche Forschungsgemeinschaft
(SFB/TR110), the Russian Science Foundation under Grant No. 16-12-10267, and the National Natural 
Science Foundation of China under Grants No. 11475181 and No. 11635009.
%\end{acknowledgments}


\end{document}